\documentclass[12pt,a4paper]{iopart}

\expandafter\let\csname equation*\endcsname=\relax
\expandafter\let\csname endequation*\endcsname=\relax

\usepackage{amsmath,amsfonts,amsthm,amssymb}
\usepackage{color}
\usepackage{graphicx}
\usepackage{hyperref}
\usepackage{subcaption}
\usepackage{tikz}
\usetikzlibrary{calc} 

\usepgflibrary{shapes.geometric}

\newcommand{\<}{\langle}
\renewcommand{\>}{\rangle}

\numberwithin{equation}{section}

\begin{document}

\title[Three-body interactions and polymer collapse]{The role of three-body interactions in two-dimensional polymer collapse\footnote{Dedicated to Professor Tony  Guttmann on the occasion of his 70th birthday}}
\author{A Bedini$^1$, A L Owczarek$^1$ and T Prellberg$^2$}
\address{$^1$ School of Mathematics and Statistics,
  The University of Melbourne, Vic 3010, Australia.}
\address{$^2$ School of Mathematical Sciences, Queen Mary University of London, Mile End Road, London E1 4NS, UK.}
\ead{abedini@unimelb.edu.au, owczarek@unimelb.edu.au, t.prellberg@qmul.ac.uk}

\date{\today}

\begin{abstract}
Various interacting lattice path models of polymer collapse in two dimensions demonstrate different critical behaviours. This difference has been without a clear explanation. The collapse transition has been variously seen to be in the Duplantier-Saleur $\theta$-point university class (specific heat cusp), the interacting trail class (specific heat divergence) or even first-order. Here we study via Monte Carlo simulation a generalisation of the Duplantier-Saleur model on the honeycomb lattice and also a generalisation of the so-called vertex-interacting self-avoiding walk model (configurations are actually restricted trails known as grooves) on the triangular lattice. Crucially for both models we have three and two body interactions explicitly and differentially weighted. We show that both models have similar phase diagrams when considered in these larger two-parameter spaces. They demonstrate regions for which the collapse transition is first-order for high three body interactions and regions where the collapse is in the Duplantier-Saleur $\theta$-point university class. We conjecture a higher order multiple critical point separating these two types of collapse.

\end{abstract}

\section{Introduction}

The nature of the collapse phase transition of a single polymer in solution when modelled in two dimensions has provided continual surprises and subtleties for the past 30 years. Connections to the rich vein of exactly solved vertex and loop models have been important in this richness. In 1987 Duplantier and Saleur \cite{duplantier1987a-a} provided a model of self-avoiding walks on the honeycomb lattice (DS model) that allowed for the conjecture of exact critical exponents for the collapse transition, also known as the $\theta$-point. After some controversy these exponents have become confirmed as those of the $\theta$-point universality class (DS class) and so pertain to the standard model of the collapse of fully-flexible lattice polymers. 

However, other models such as a Bl\"ote-Nienhuis model \cite{blote1989a-a}, the so-called vertex interaction self-avoiding walk (VISAW) model \cite{foster2003a-a,foster2003b-a,foster2011a-a,bedini2013c-:a}  as well as interacting lattice trails \cite{shapir1984a-a,owczarek1995a-:a,owczarek2006a-:a} seem to demonstrate different collapse behaviour. Our usual understanding from the principle of universality is that minor changes in microscopic details should not affect the university class. For the collapse transition this would seem to be called into question.  More generally, for collapse models, the addition of stiffness, or higher order interactions, can result in different low temperature phases, and even first-order collapse. For example,  Doukas \emph{et al.\ }\cite{doukas2010a-:a} studied a model of interacting trails on the triangular lattice, where doubly and triply visited sites can be weighted differently, and found a richer phase diagram with the DS collapse transition changing over to a first-order transition depending on the parameters. The question naturally arises as to whether such a scenario can be seen in models without the topological complication of crossing paths.

In this paper we study a generalisation of the DS model \cite{duplantier1987a-a} where we differentially weight faces of `\emph{type-}2' and `\emph{type-}3' which relate to faces of the honeycomb lattice visited by two and three separate parts of the walk respectively. We also study a generalisation of the  VISAW model on the triangular lattice similar to the model studied by Doukas \emph{et al.\ }\cite{doukas2010a-:a}. We find that in both cases a similar phase diagram eventuates --- one we conjecture to be the same as that described by Doukas \emph{et al.\ }\cite{doukas2010a-:a}. This would indicate that trails, and the subset of non-crossing trails called grooves found in the VISAW, and also canonical self-avoiding walk models of polymer collapse actually all have similar phase diagrams and university classes when viewed in the space of two and three body interactions.

\section{The models}

Let us first define the two models in which we are interested more precisely.

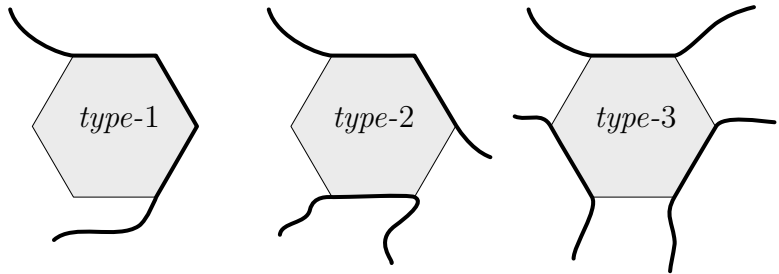
\begin{figure}[t]
	\centering
	\begin{tikzpicture}
	\node (fig1) at (0,0) {\includegraphics[scale=0.75]{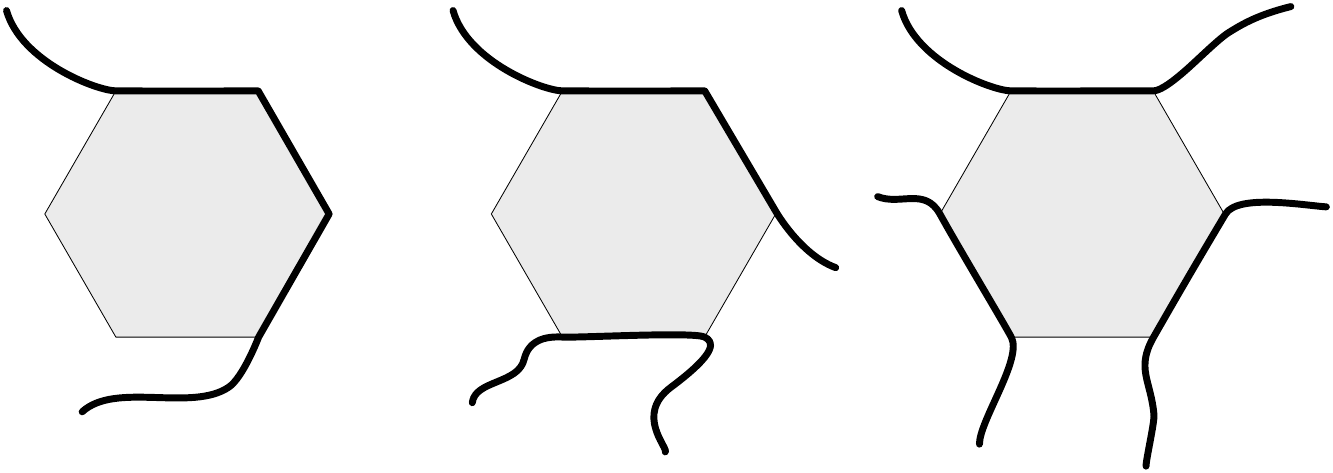}};
	\node (f1) at (-3.60cm,0.25cm) {\emph{type-}$1$};
	\node (f2) at (-0.25cm,0.25cm) {\emph{type-}$2$};
	\node (f3) at (3.20cm,0.25cm) {\emph{type-}$3$};
	\end{tikzpicture}
	\caption{In the DS model faces of the underlying hexagonal lattice are classified by how many distinct segments of the the walk are in contact with the face. In this picture we illustrate an example of each category.}
	\label{fig:hex-faces}
\end{figure}

\noindent\textbf{Generalised DS model}. Consider the ensemble $\mathcal S_n$ of self-avoiding lattice walks of $n$ steps that can be formed on the hexagonal lattice.
Given a SAW $\varphi_n \in \mathcal S_n$, we highlight every face that $\varphi_n$ touches and we divide them in three categories depending on how many distinct segments of the walk are in contact with the face. We then count the number of faces belonging to each category obtaining the numbers $f_i$ for $i \in \{1,2,3\}$. Examples of each face category are illustrated in Figure~\ref{fig:hex-faces}. We shall refer to them as type-$i$ faces respectively.

Introducing weights $\omega_2$ and $\omega_3$ dual to $f_2$ and $f_3$ we define the partition function for this model,
\begin{equation}
	Z_n(\omega_2, \omega_3) = \sum_{\varphi_n\in\mathcal S_n}\ \omega_2^{f_2} \omega_3^{f_3}.
\end{equation}
The model described by Duplantier and Saleur \cite{duplantier1987a-a}  has $\omega_2=2$ and $\omega_3=\omega_2^2=4$.

As usual, the finite-length reduced free energy is then given by
\begin{equation}
  \kappa_n(\omega_2, \omega_3) = \frac{1}{n} \log\, Z_n(\omega_2, \omega_3),
  \label{eq:free-energy-1}
\end{equation}
and the thermodynamic limit is obtained by taking the limit of large $n$, i.e.,
\begin{equation}
  \kappa(\omega_2, \omega_3) = \lim_{n \to \infty} \kappa_n(\omega_2, \omega_3).
  \label{eq:free-energy-2}
\end{equation}
Thermodynamic quantities like the internal energy and the specific heat are obtained by taking derivatives of the free energy,
\begin{equation}
	u^{(i)}_n = \frac{\partial \kappa_n}{\partial \omega_i},
	\qquad
	c^{(i)}_n = \frac{\partial^2 \kappa_n}{\partial \omega_i^2}.
\end{equation}
For some values of $\omega_2$ and $\omega_3$ we expect the thermodynamic limit to be singular, that is, function such as $c^{(i)}(\omega_2,\omega_3)$ will be non-analytic on some manifold and behave as
\begin{equation}
	c^{(i)} = \lim_{n \to \infty} c^{(i)}_n \sim B\, |t|^{-\alpha},
\end{equation}
where $t$ is an appropriate distance from the critical manifold. For example, we know that the DS point $(\omega_2,\omega_3)=(2,4)$ is critical.

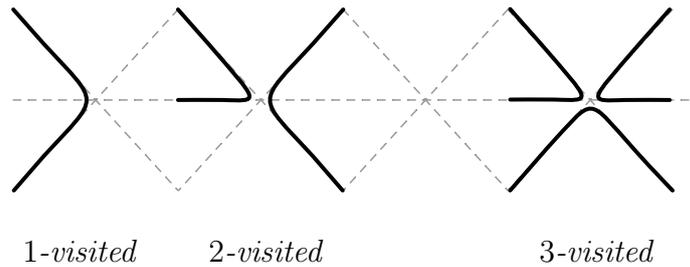
\begin{figure}[t]
	\centering
	\begin{tikzpicture}
	\node (fig1) at (0,0) {\includegraphics[scale=0.75]{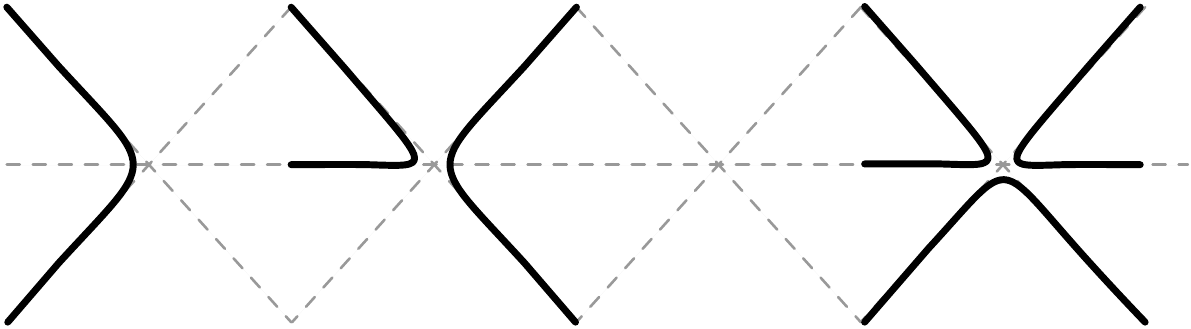}};
	\node (f1) at (-3.60cm,-2cm) {$1$\emph{-visited}};
	\node (f2) at (-1.15cm,-2cm) {$2$\emph{-visited}};
	\node (f3) at (3.20cm,-2cm) {$3$\emph{-visited}};
	\end{tikzpicture}
	\caption{In the interacting groove (IG) model sites of the triangular lattice are classified by how many distinct segments of the groove are incident on a site. In this picture we illustrate an example of each category.}
	\label{fig:tri-faces}
\end{figure}

\noindent\textbf{Interacting Grooves (IG)}. For the second model we consider the set $\mathcal G_n$ of \emph{grooves} on the triangular lattice. This is the set of bond-avoiding walks (trails) with the added restriction that no crossings are allowed. This is an important type of configuration since  these configurations appear in the high temperature expansion of the $O(n)$ model on the square lattice (see for example the recent work in \cite{fu2013a-a}). For each configuration we count the number of sites that the walk visits two and three times --- see Figure~\ref{fig:tri-faces}. Let these numbers be $m_2$ and $m_3$ and refer to the sites as $i$-visited, or rather doubly-visited and triply-visited. We then introduce weights $\tau_1$ and $\tau_2$ and we define, in complete analogy with above, the following quantities:
\begin{eqnarray}
	Z_n &= \sum_{\varphi_n\in\mathcal G_n}\ \tau_2^{m_2} \tau_3^{m_3}, \\
	\kappa_n &= \frac{1}{n} \log\, Z_n(\tau_2, \tau_3), \\
	u^{(i)}_n &= \frac{\partial \kappa_n}{\partial \tau_i}, \\
	c^{(i)}_n &= \frac{\partial^2 \kappa_n}{\partial \tau_i^2}.	
\end{eqnarray}
We shall refer to this model as the Interacting Groove (IG) model.
\section{Simulations}

We studied the two models using the FlatPERM algorithm \cite{prellberg2004a-a} which is based on the Pruned and Enriched Rosenbluth Method (PERM) developed in \cite{grassberger1997a-a}.

For the PERM algorithm, at each iteration a polymer configuration is generated kinetically (which is to say that each growth step is selected at random from all possible growth steps) along with a weight factor to correct the sample bias. At each growth step, configurations with very high weight relative to other configurations of the same size  are enriched (duplicated) while configurations with low weight or that cannot be grown any further are pruned (discarded). Despite introducing a correlation between each iteration, this simple mechanism greatly improves the algorithm efficiency. A single iteration is then concluded when all configurations have been pruned and the total number of samples generated during each iteration depends on the problem at hand and on the details of the enriching/pruning strategy.

FlatPERM extends this method by cleverly choosing the enrichment and pruning steps to generate for each polymer size $n$ a quasi-flat histogram in some chosen micro-canonical quantities $\mathbf{k}=(k_1,k_2,\dotsc,k_{\ell})$ and producing an estimate $W_{n,\mathbf{k}}$ of the total weight of the walks of length $n$ at fixed values of $\mathbf{k}$. From the total weight one can access physical quantities over a broad range of temperatures through a simple weighted average, e.g.
\begin{align}
  \< \mathcal O \>_n(\rho) = \frac{\sum_{\mathbf{k}} \mathcal O_{n,\mathbf{k}}\,
   \left(\prod_j \rho_j^{k_j}\right) \, W_{n,\mathbf{k}}}{\sum_\mathbf{k} \left(\prod_j \rho_j^{k_j}\right) \, W_{n,\mathbf{k}}}.
\end{align}
The quantities $k_j$ may be any subset of the physical parameters of the model. To study the full two parameter phase space of the generalised DS model one would set $(k_1, k_2) = (f_2, f_3)$ and $(\rho_1, \rho_2) = (\omega_2, \omega_3)$. Thermodynamic quantities are then calculated by
\begin{equation}
	u_n^{(i)} = \frac{\sum_{f_2,f_3} f_i \; \omega_2^{f_2} \omega_3^{f_3} \; W_{n,f_2,f_3}}{\sum_{f_2,f_3}\omega_2^{f_2} \omega_3^{f_3} \; W_{n,f_2,f_3}}.
\end{equation}

\subsection{Three-body DS model}

We have run the flatPERM algorithm limiting the configurations to a length of $n = 256$ and running $7.32 \cdot 10^6$ iterations. In this way we produced $5.72 \cdot 10^{10}$ samples at the maximum length. As is usual with the flatPERM algorithm, one can also count the configurations by the fraction of its steps made independently, this gives a measure of the number of ``effectively independent samples''. Our numerical study collected $2.79 \cdot 10^9$ effective samples.

\subsection{Three-body interacting grooves}

We have run the flatPERM algorithm limiting the configurations to a length of $n = 256$ and running $7.91 \cdot 10^5$ iterations. In this way we produced $1.1 \cdot 10^{10}$ samples at the maximum length. As common with the flatPERM algorithm, one can also count the configurations by the fraction of its steps made independently, this gives a measure of the number of ``effectively independent samples''. Our numerical study collected $2.4 \cdot 10^8$ effective samples.

\section{Results}

\subsection{Three-body DS model}

Before considering the larger model note that when $\omega_2=\omega_3=1$ we simply have self-avoiding walks on the honeycomb lattice, which is a well studied model. Non-interacting self-avoiding walks behave as extended geometric objects relative to random walks and are dominated by the so-called excluded volume effect. The long standing conjecture is that the exponent $\nu$ describing the growth of the radius of gyration equals $3/4$ \cite{nienhuis1982a-a}. Here we would expect the ``extended" behaviour to exist in a region around the point $(1,1)$ in the larger parameter space.

\begin{figure}[ht!]
\begin{center}
\includegraphics[width=0.8\columnwidth]{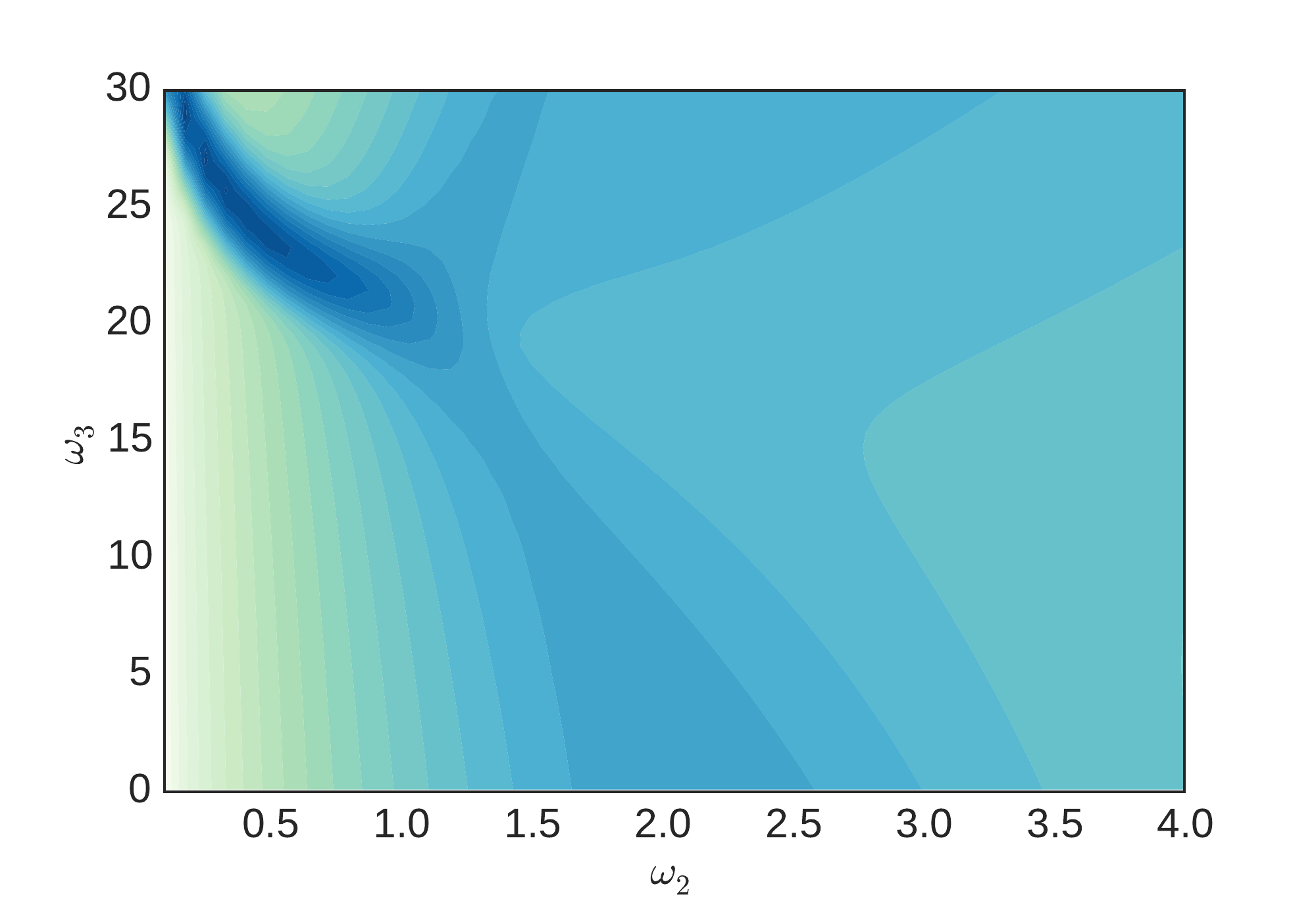}
\caption{Density plot of the logarithm of the largest eigenvalue of the matrix of second derivatives of the free energy with respect to $\omega_2$ and $\omega_3$ at length 256. Darker shades (colours) represent larger values. There is a clear indication of a strong transition for large $\omega_3$ in the range 20 to 30 when $\omega_2$ is less than $1.5$. %
}\label{DS-hessian}
\end{center}
\end{figure}

\begin{figure}[ht!]
\begin{center}
\includegraphics[width=0.6\columnwidth]{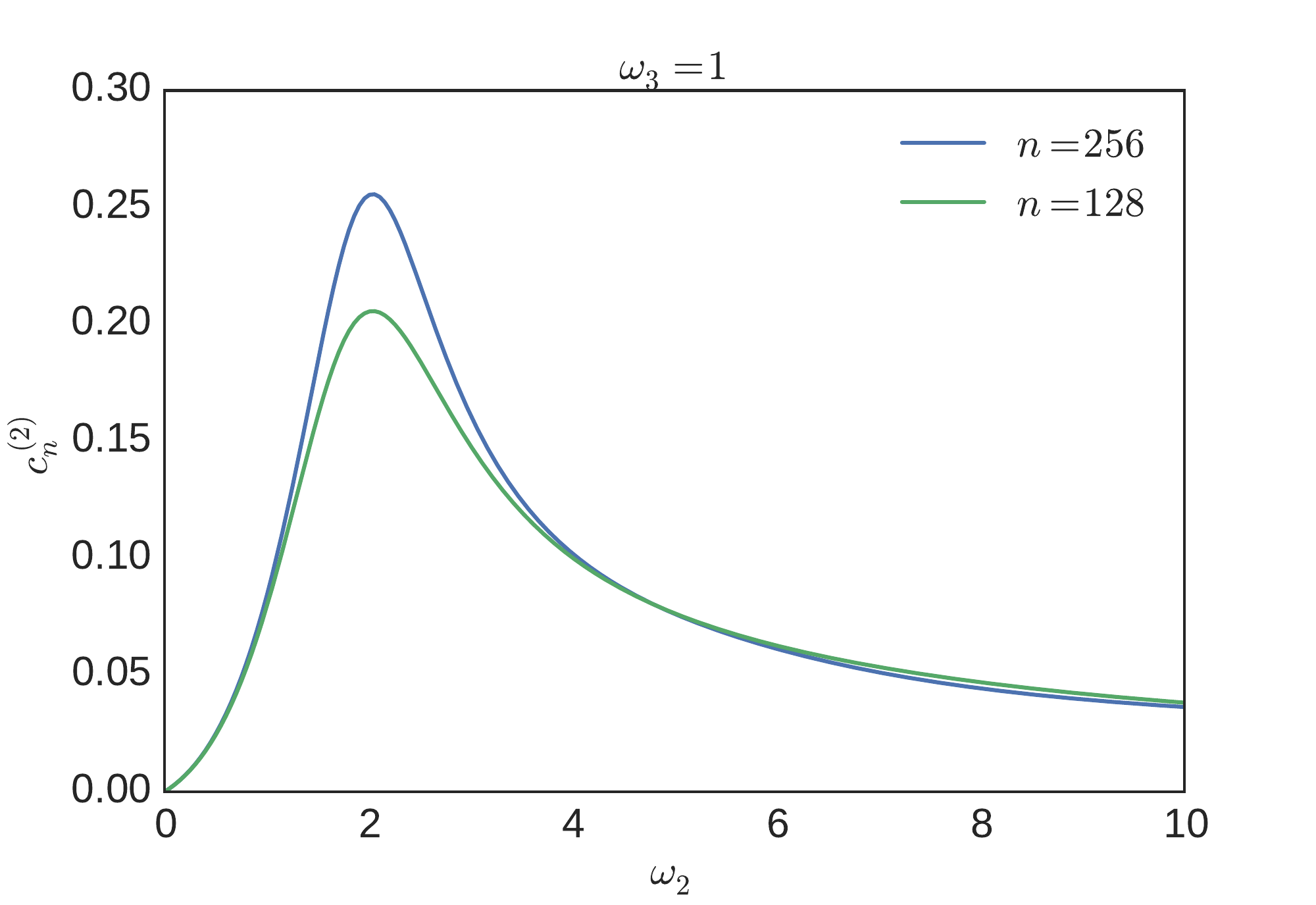}
\caption{Plot of the specific heat as a function of $\omega_2$ for lengths $n=128$ and $n=256$. In this plot $\omega_3$ is fixed to 1. The peak of the specific heat has only grown slowly between these lengths.%
}\label{DS-spec-heat-2}
\end{center}
\end{figure}

\begin{figure}[ht!]
\begin{center}
\includegraphics[width=0.6\columnwidth]{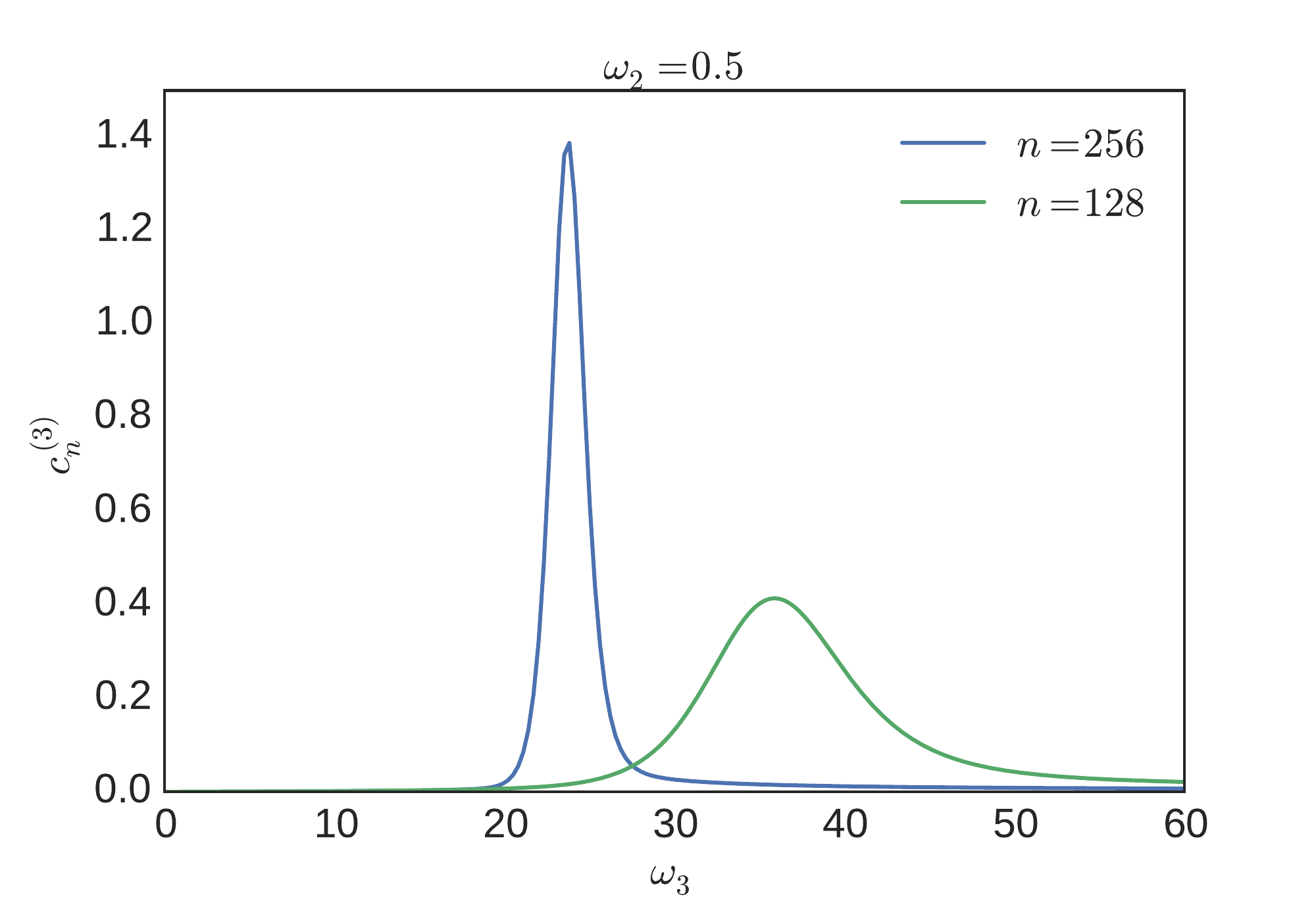}
\caption{Plot of the specific heat as a function of $\omega_3$  for lengths $n=128$ and $n=256$. In this plot $\omega_2$ is fixed to 0.5. There has been a very dramatic change between these two lengths with the maximum value more than doubling.%
}\label{DS-spec-heat-3}
\end{center}
\end{figure}

\begin{figure}[ht!]
\begin{center}
\includegraphics[width=0.6\columnwidth]{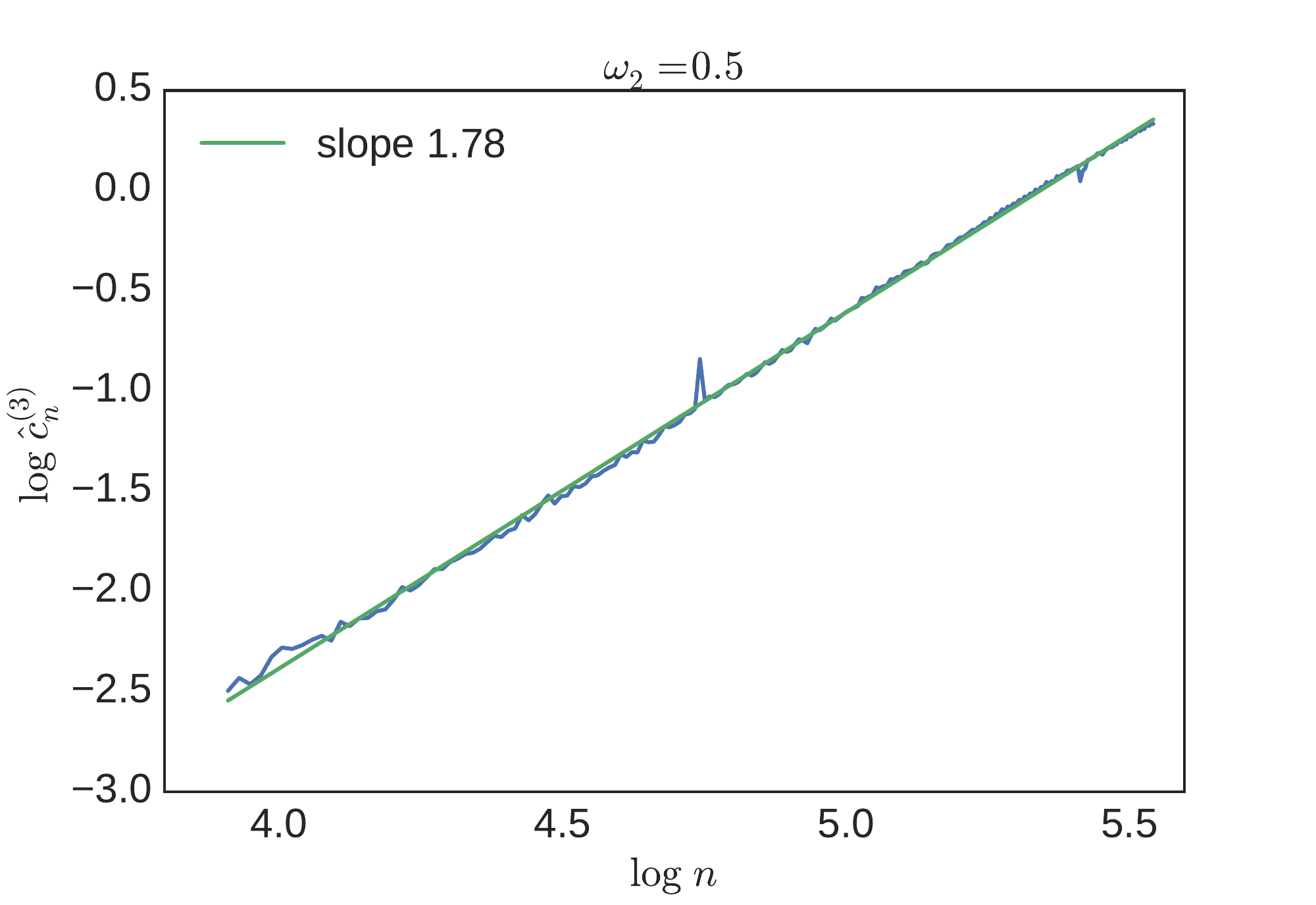}
\caption{Plot of the logarithm of the maximum value of the specific heat against $\log (n)$.  A local exponent fit over the short range of lengths  (50  to 256) considered here gives a value of $\alpha\phi \approx 1.8$. The theoretical maximum asymptotic scaling would give an exponent of one: clearly large corrections to scaling are still at play here. In this plot $\omega_2$ is fixed to 0.5.%
}\label{DS-spec-heat-diverge}
\end{center}
\end{figure}

\begin{figure}[ht!]
\begin{center}
\includegraphics[width=0.6\columnwidth]{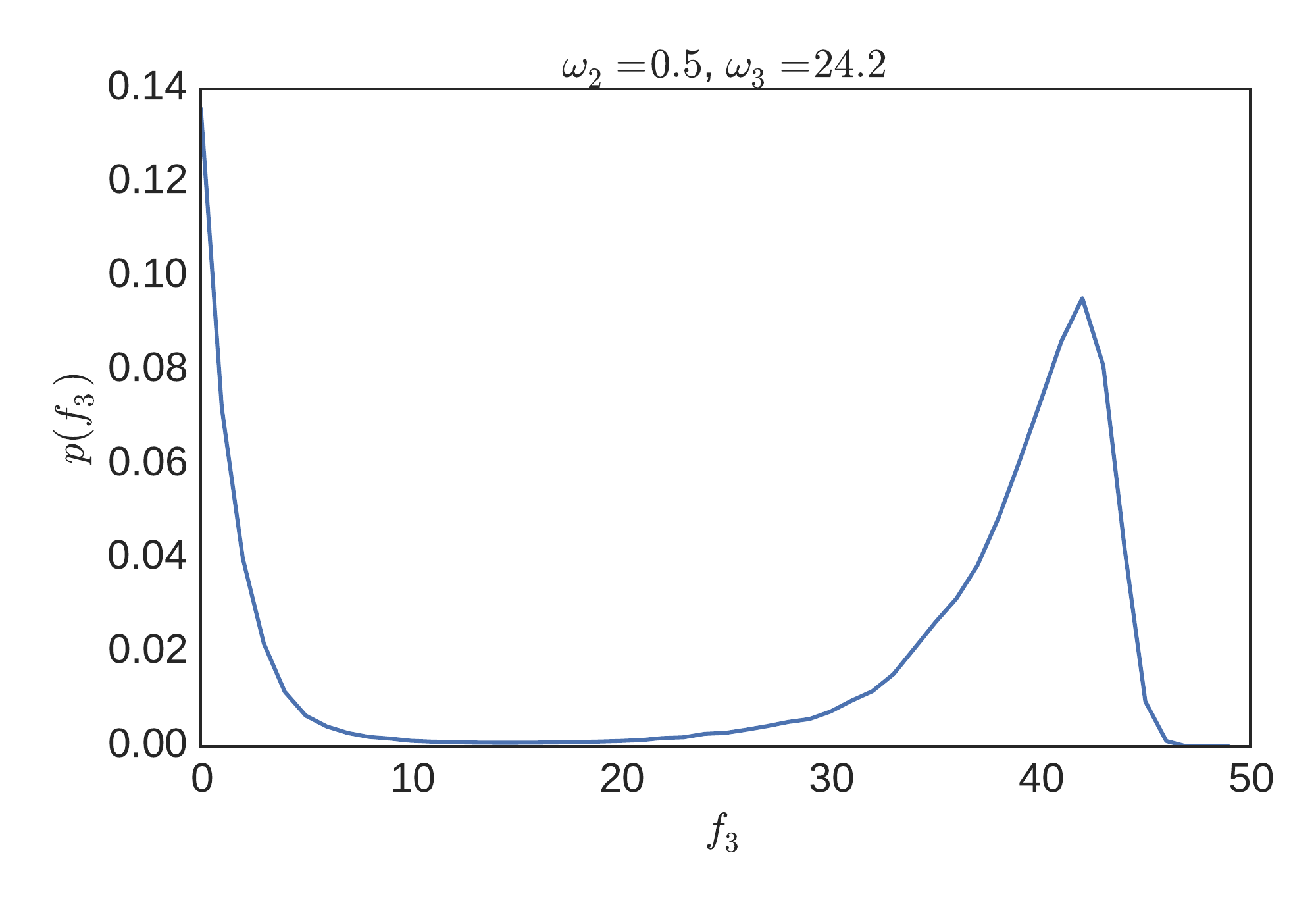}
\caption{The distribution of the number of \emph{type}-3 faces $f_3$ is clearly bimodal at the point when $\omega_2$ and $\omega_3$ cross the line of suspected first-order transitions.%
}\label{DS-double-peak}
\end{center}
\end{figure}

\begin{figure}[ht!]
\begin{center}
\includegraphics[width=\columnwidth]{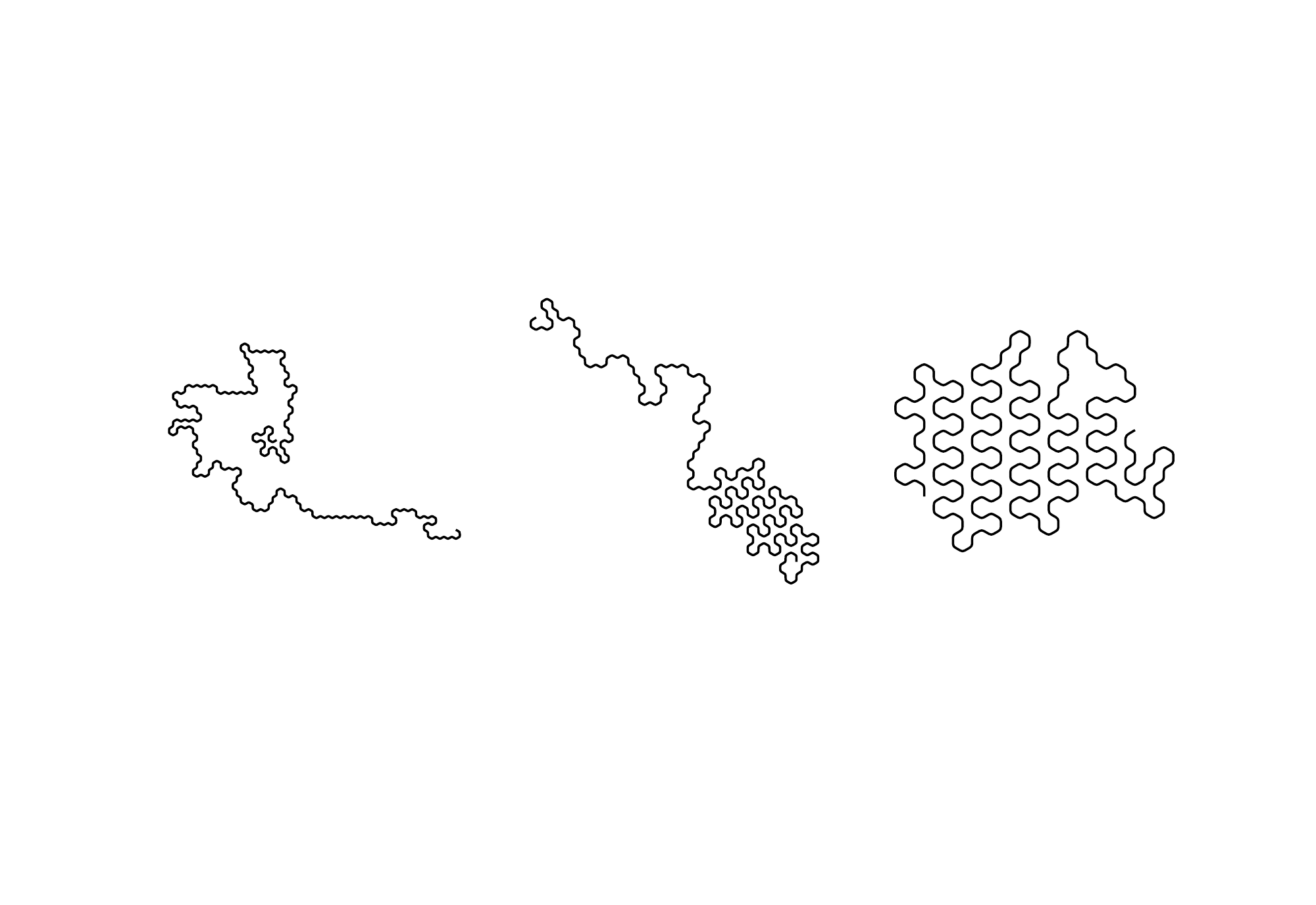}
\vspace{-18ex}\caption{Typical configurations of walks that have been generated at $\omega_2 = 0.5$ with different values of $\omega_3$. From left to right: $\omega_3$ = 22, $\omega_3$ = 24, $\omega_3$ = 26. The configurations illustrate the co-existence of fully dense and swollen parts of the polymer, demonstrating the first-order nature of the transition.%
}\label{DS-typical-configs}
\end{center}
\end{figure}

To obtain a landscape of possible phase transitions, we plot the largest eigenvalue of the matrix of second derivatives of the free energy with respect to $\omega_2$ and $\omega_3$ (measuring the strength of the fluctuations and covariance in $f_2$ and $f_3$) at length $n=256$ in Figure~\ref{DS-hessian}. 
The original DS model with $\omega_3=\omega_2^2$ has the weak signature of the $\theta$-point around the exact point at $\omega_2=2$ where we expect a cusp singularity (the third derivative of the free energy is divergent).
Considering the slice $\omega_3=1$ we plot the specific heat  in Figure~\ref{DS-spec-heat-2}, which shows almost no sign of a transition at all. This is expected with $\alpha=-1/3$ at a $\theta$-point. Right at this point $\nu=4/7$ \cite{duplantier1987a-a}.
It is striking though that for $\omega_2<1.5$ and $\omega_3$ in the range $20 < \omega_3 < 30$ the fluctuations are quite large indicating the build up of a strong transition. 
In Figure~\ref{DS-spec-heat-3} we first plot the specific heat for two lengths: the enormous difference between the graphs for the two lengths points to this build up.
We then find the value of the maximum of the specific heat for a range of lengths and plot these on a log-log plot in Figure~\ref{DS-spec-heat-diverge}. The build up is so strong that is is super-linear: asymptotically the build up can only be at worst linear. It is usual to interpret a super linear build up as a linear asymptotic regime with strong corrections to scaling. Now, such a linear build up asymptotically indicates a first-order transition thermodynamically. 
To test the hypothesis of a first-order transition we consider the distribution of \emph{type}-3 faces at the location of the maximum specific heat for length $n=256$ in Figure~\ref{DS-double-peak}: a clear double peak distribution emerges which is a signature of a first-order transition build up.
To confirm the type of transition we illustrate typical configurations in Figure~\ref{DS-typical-configs} generated at $\omega_2=0.5$ for three values of $\omega_3$ above, at, and below the finite size transition location, as inferred from the location of the specific heat maximum.
The configuration found at high $\omega_3=26$ looks like an ordered and anisotropic crystal-like dense structure while that at the smallest value of $\omega_3=22$ looks very extended as one would expect of a self-avoiding walks without interactions. The configuration collected at the peak of the specific heat, located at $\omega_3=24$, seems to be a simple phase separation of the extended high temperature type of configuration and the compact crystal. This reinforces the conclusion that we have identified a first-order transition. 

\begin{figure}[ht!]
\begin{center}
\includegraphics[width=0.6\columnwidth]{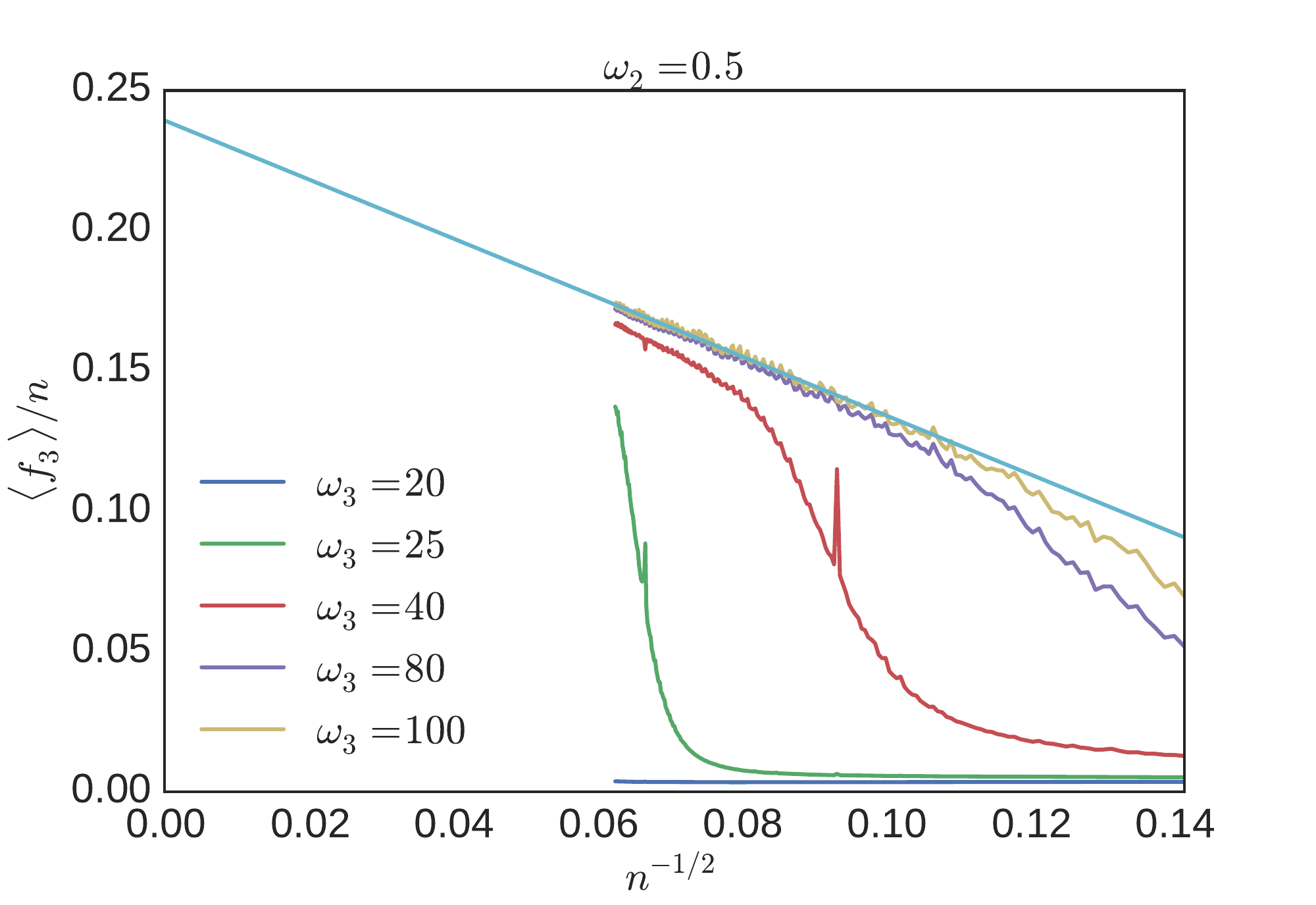}
\caption{Plot of $\<f_3\>/n$ as a function of $n^{-1/2}$ for various values of $\omega_3$. In this plot $\omega_2$ is fixed to 0.5. The line is the graph of the linear extrapolation for the largest $\omega_3$ value considered. For $\omega_3 \geq 40$ we estimate the $\lim_{n\rightarrow\infty} \<f_3\>/n$ to be slightly less than $1/4$, though not unambiguously so.%
}\label{DS-low-temp-order}
\end{center}
\end{figure}

Turning to the low temperature phase itself we consider the configuration generated at $\omega_3=26$. Note that apart from near the boundaries the faces alternate between \emph{type-}3  and \emph{type-}1 without any \emph{type-}2 faces. This would lead to half the faces being \emph{type-}3 in the thermodynamic limit. As each step is associated with two faces the limiting density of \emph{type-}3 and \emph{type-}1 faces should be $1/4$ in this scenario. So to further understand the low temperature phase we plot in Figure~\ref{DS-low-temp-order} the average density of \emph{type-}3 faces at a range of $\omega_3$ moving from moderate to larger values deep into the low temperature regime.
We see that for large enough $\omega_3$ the density of \emph{type-}3 faces scales as
\begin{equation}
\<f_3\>/n \sim F  - \frac{c}{n^{1/2}}
\end{equation}
 characteristic of a low temperature dense phase. If the low-temperature phase is fully dense then $F=1/4$ throughout the low-temperature phase for any $\omega_3$ larger than the transition value. At the lengths considered we estimate $F$ to be slightly less than $1/4$, though not unambiguously so. We rather conjecture though that the low temperature regime is indeed fully dense for large $\omega_3$ and small $\omega_2$. On the other hand for large $\omega_2$ when $\omega_3$ is moderate in value we find that $F \ll 1/4$, pointing to the presence of the expected globular phase of low temperature ISAW.
 
Returning to Figure~\ref{DS-hessian} one can make out three possible phase boundaries that may meet at a point. One of those boundaries we would conjecture is a line of first-order transitions found when varying $\omega_3$ for small $\omega_2$, while another is a line of DS $\theta$-point like transitions with $\alpha=-1/3$ seen when varying $\omega_2$ for $\omega_3$ up to perhaps 20. There does also seem to be a line of low temperature transitions separating the two phases we have identified above. That is, this would be a transition between the liquid-drop-like amorphous globule for large $\omega_2$ and the very dense crystal-like phase we find here for large $\omega_3$.
 
\begin{figure}[ht!]
\begin{center}
\includegraphics[width=0.8\columnwidth]{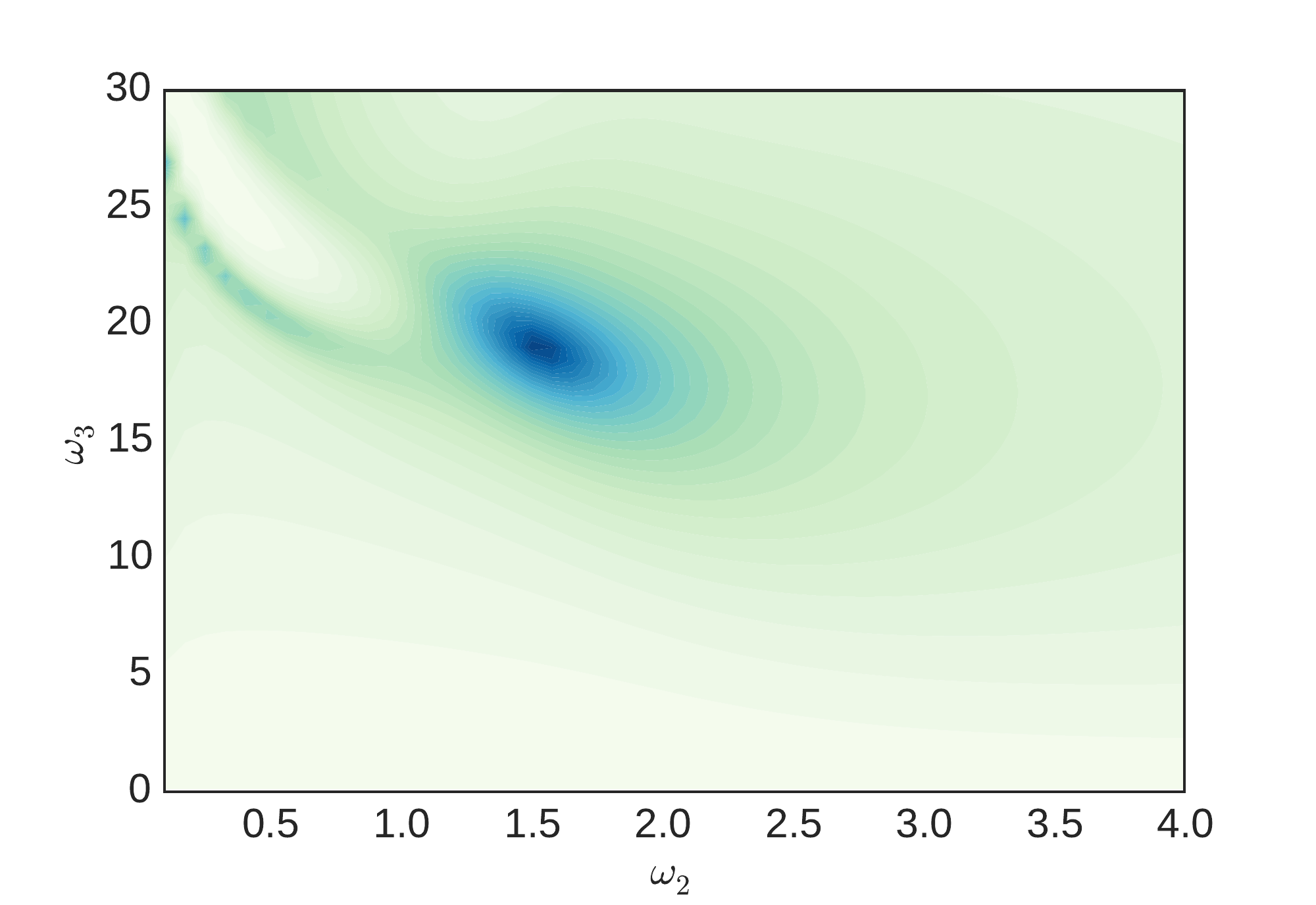}
\caption{Density plot of the ratio ${\lambda}_{min}/{\lambda}_{max}$ of the eigenvalues of the matrix of second derivatives of the free energy with respect to $\omega_2$ and $\omega_3$ at length 256. Darker shades (colours) represent larger values. There is a clear indication of a special point near $\omega_2=1.5$ and $\omega_3=20$.%
}\label{DS-hessian-ratio}
\end{center}
\end{figure}

 If this scenario described is correct these three phase boundaries should meet at a  point. In previous work it was noticed that an accurate way to find the location of such a multi-critical point is to look at the ratio of the maximum and minimum eigenvalues of the matrix of second  derivatives of the free energy with respect to the parameters of the model. We make such a plot here in Figure~\ref{DS-hessian-ratio}.
We see that this plot does pick out a particular location $(\omega_2,\omega_3)=(1.5,20)$ that seems to be at the intersection of the three finite size phase boundaries.

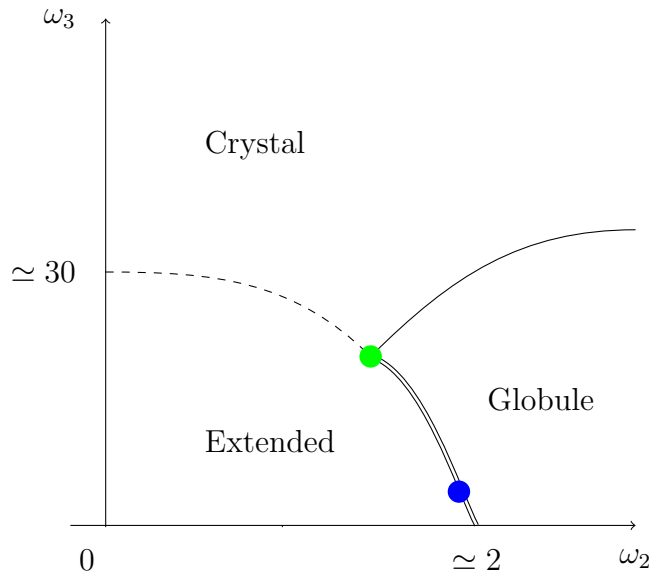
\begin{figure}[ht!]
\begin{center}
\begin{tikzpicture}[x=1.66cm,y=0.08cm,scale=1.4]
  \coordinate (A) at (2.1,0);
  \coordinate (C) at (1.5,20);
  \coordinate (B) at (0,30);
  \coordinate (DS) at (2,4);
  \coordinate (C1) at ($(C)!0.5cm!40:(DS)$);
  
  \draw[double,double distance=1pt] (A) .. controls (DS) and (C1) .. (C);
  
   \draw[dashed] (0,30) to [out=0, in =135] (1.5,20);
   
  \draw[solid] (1.5,20) to [out=45, in =180] (3,35);
  \fill[blue] (2,4) circle (3pt);
  
  \draw (0.5,10) node[align=left, right]{Extended};
   \draw (2.1,15) node[align=left, right]{Globule};
   \draw (0.5,45) node[align=left, right]{Crystal};

  \draw[->] (-0.2,0)
  	-- (0,0) node[left=7pt,below=5pt] {0}
    -- (A) node[below=5pt] {$\simeq 2$}
    -- (3,0) node[below=5pt] {$\omega_2$};
    
  \draw[->] (0,-0.2)
  	-- (B) node[left=7pt] {$\simeq 30$}
    -- (0,60) node[left=7pt] {$\omega_3$};

  \draw[black,solid] (1,-0.2) -- (1,0) ;

  \fill[green] (C) circle (3pt);
\end{tikzpicture}
\caption{Schematic phase diagram for the generalised DS model involving self-avoiding walks on the honeycomb lattice. There are three phases: Extended, Globule and Crystal. The line of DS second-order phase transitions between the Extended and Globule phases is marked with double lines. The first-order line between the extended and Crystal phases is marked as a dashed line. The meeting of these is marked by a bullet (green) at $(\omega_2,\omega_3)=(1.5,20)$. We also mark with a bullet (blue) the DS critical point at $(\omega_2,\omega_3)=(2,4)$. We mark the possible phase transition between two types of low temperature phase, Globule and Crystal, with a single solid line, indicating a probable second-order transition.}\label{DS-phase-diagram}
\end{center}
\end{figure}

Putting all this information together we conjecture a phase diagram schematically shown in Figure~\ref{DS-phase-diagram}.
We have three phase boundaries meeting at a point marked with a bullet separating three phases marked as extended, globule and crystal. The phase boundary between the extended and globule phase we conjecture will be uniformly in the universality class described by Duplantier and Saleur \cite{duplantier1987a-a}: we see little difference between setting $\omega_3=\omega_2^2$ as Duplantier and Saleur \cite{duplantier1987a-a} did in their model and simply setting $\omega_3=1$. The phase boundary between the extended and crystal phase is very strong, and overhelming evidence suggests a first-order transition. We have not investigated the globule-crystal transition in this work and certainly that would be of real interest in future work.

\subsection{Three-body IG model}
\begin{figure}[ht!]
\begin{center}
\includegraphics[width=0.8\columnwidth]{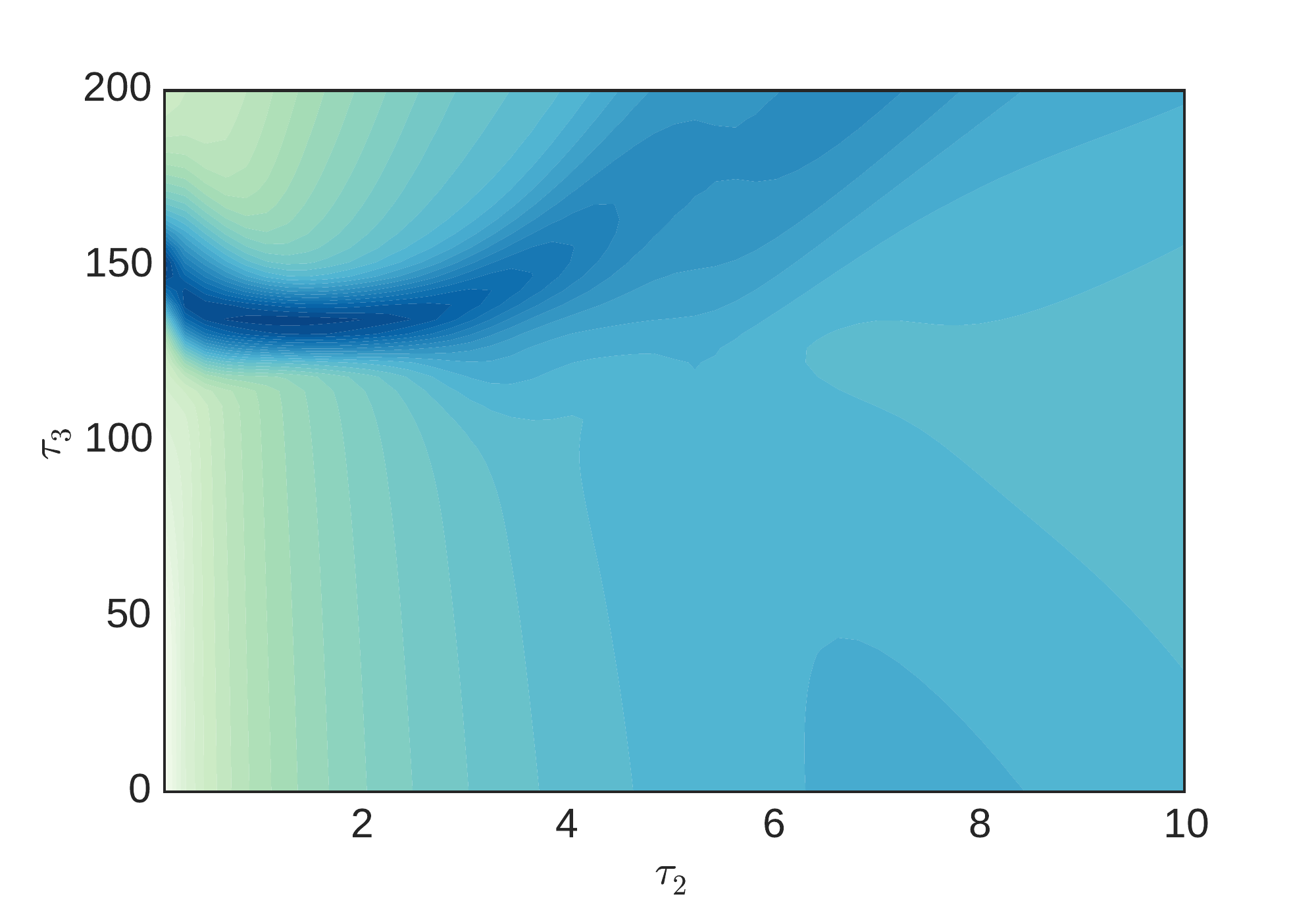}
\caption{Density plot of the logarithm of the largest eigenvalue of the matrix of second derivatives of the free energy with respect to $\tau_2$ and $\tau_3$ at length 256. Darker shades (colours) represent larger values. There is a clear indication of a strong transition for large $\tau_3$ in the range 120 to 160 when $\omega_2$ is less than $4$. %
}\label{IG-hessian}
\end{center}
\end{figure}

\begin{figure}[ht!]
\begin{center}
\includegraphics[width=0.8\columnwidth]{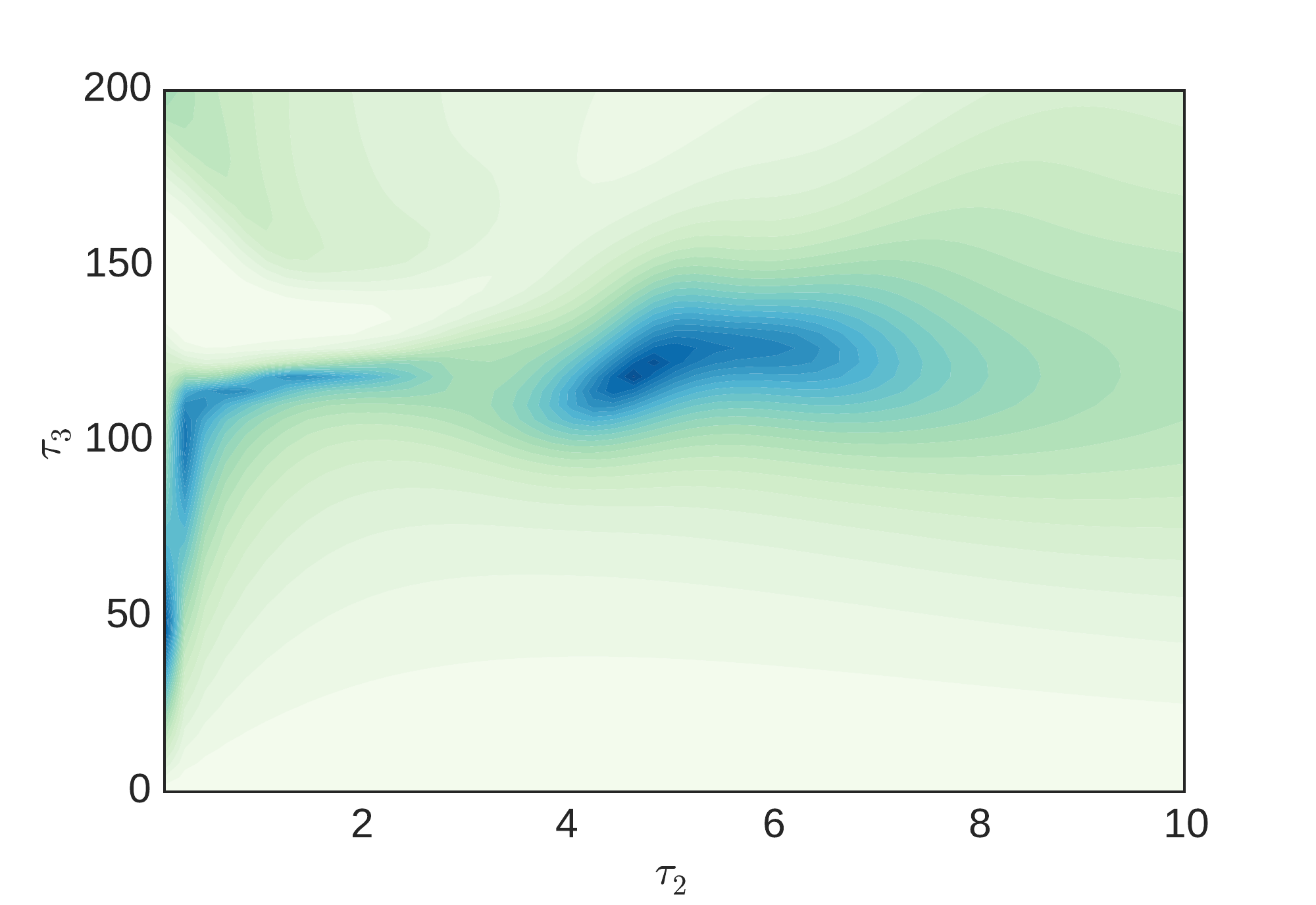}
\caption{Density plot of the ratio ${\lambda}_{min}/{\lambda}_{max}$ of the eigenvalues of the matrix of second derivatives of the free energy with respect to $\tau_2$ and $\tau_3$ at length 256. Darker shades (colours) represent larger values. There is an indication of a special point near $\omega_2=4.5$ and $\omega_3=125$.%
}\label{IG-hessian-ratio}
\end{center}
\end{figure}

\begin{figure}[ht!]
\begin{center}
\includegraphics[width=0.6\columnwidth]{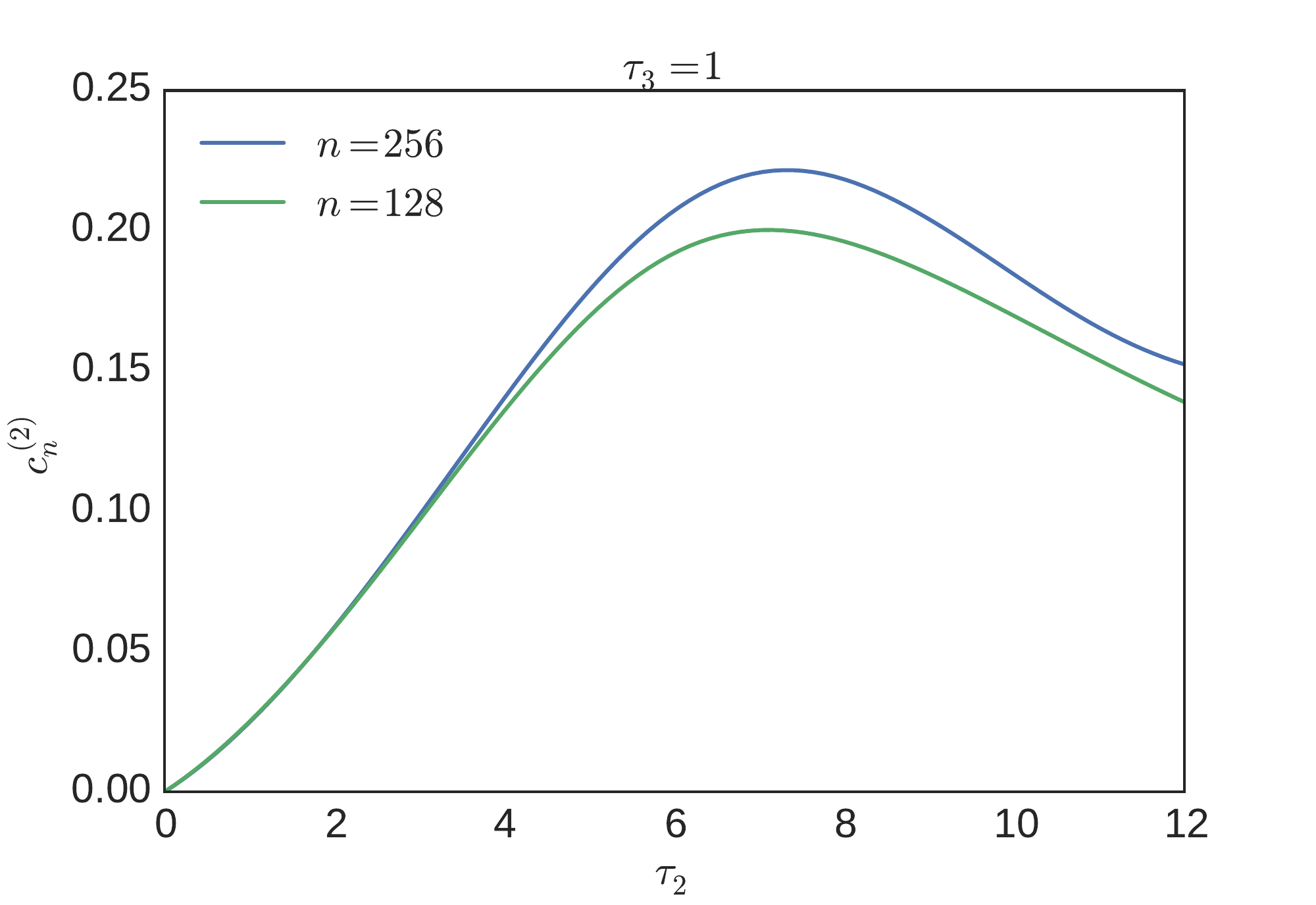}
\caption{Plot of the specific heat as a function of $\tau_2$, in this plot $\tau_3$ is fixed to 1. The peak of the specific heat has only grown slowly between these lengths.%
}\label{IG-spec-heat-2}
\end{center}
\end{figure}

\begin{figure}[ht!]
\begin{center}
\includegraphics[width=0.6\columnwidth]{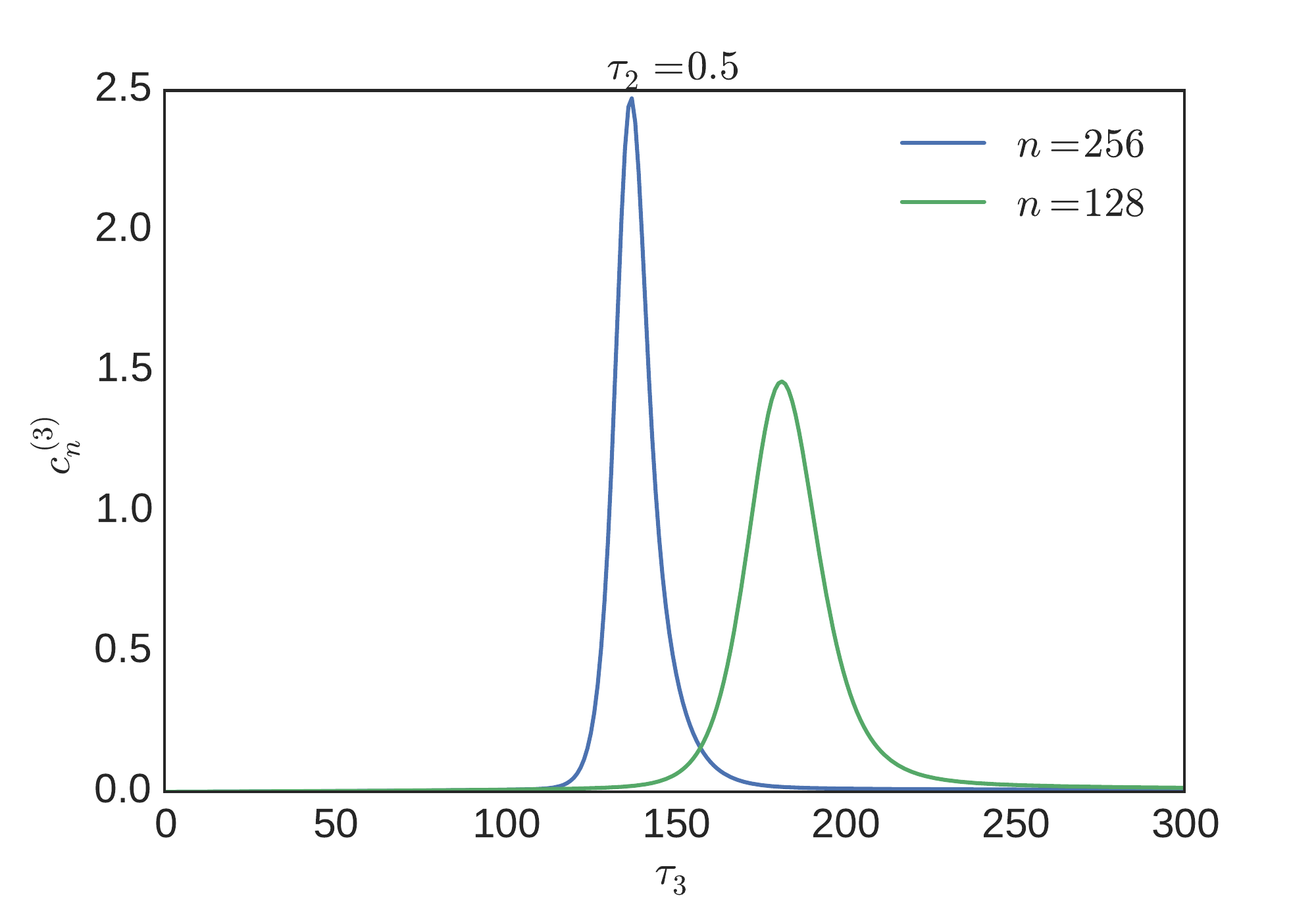}
\caption{Plot of the specific heat as a function of $\tau_3$, in this plot $\tau_2$ is fixed to 0.5. There has been a very dramatic change between these two lengths with the maximum value close to doubling.%
}\label{IG-spec-heat-3}
\end{center}
\end{figure}

\begin{figure}[ht!]
\begin{center}
\includegraphics[width=0.6\columnwidth]{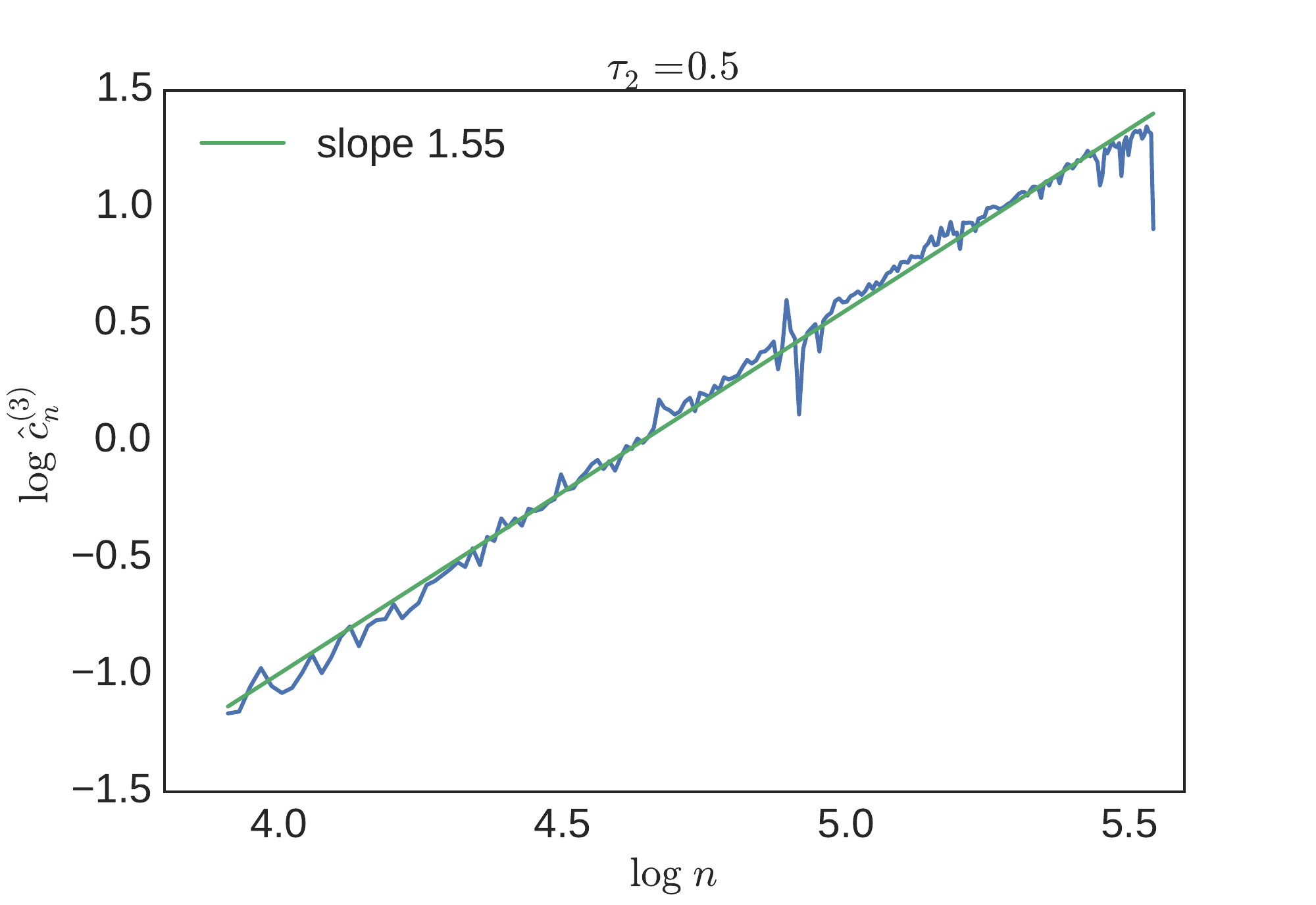}
\caption{Plot of the maximum value of the specific heat as a function of $\tau_3$. A local exponent fit over the short range of lengths  (50  to 256) considered here gives a value of $\alpha\phi \approx 1.5$. The theoretical maximum asymptotic scaling would give an exponent of one: clearly large corrections to scaling are still at play here. In this plot $\tau_2$ is fixed to 0.5.%
}\label{IG-spec-heat-diverge}
\end{center}
\end{figure}

\begin{figure}[ht!]
\begin{center}
\includegraphics[width=0.6\columnwidth]{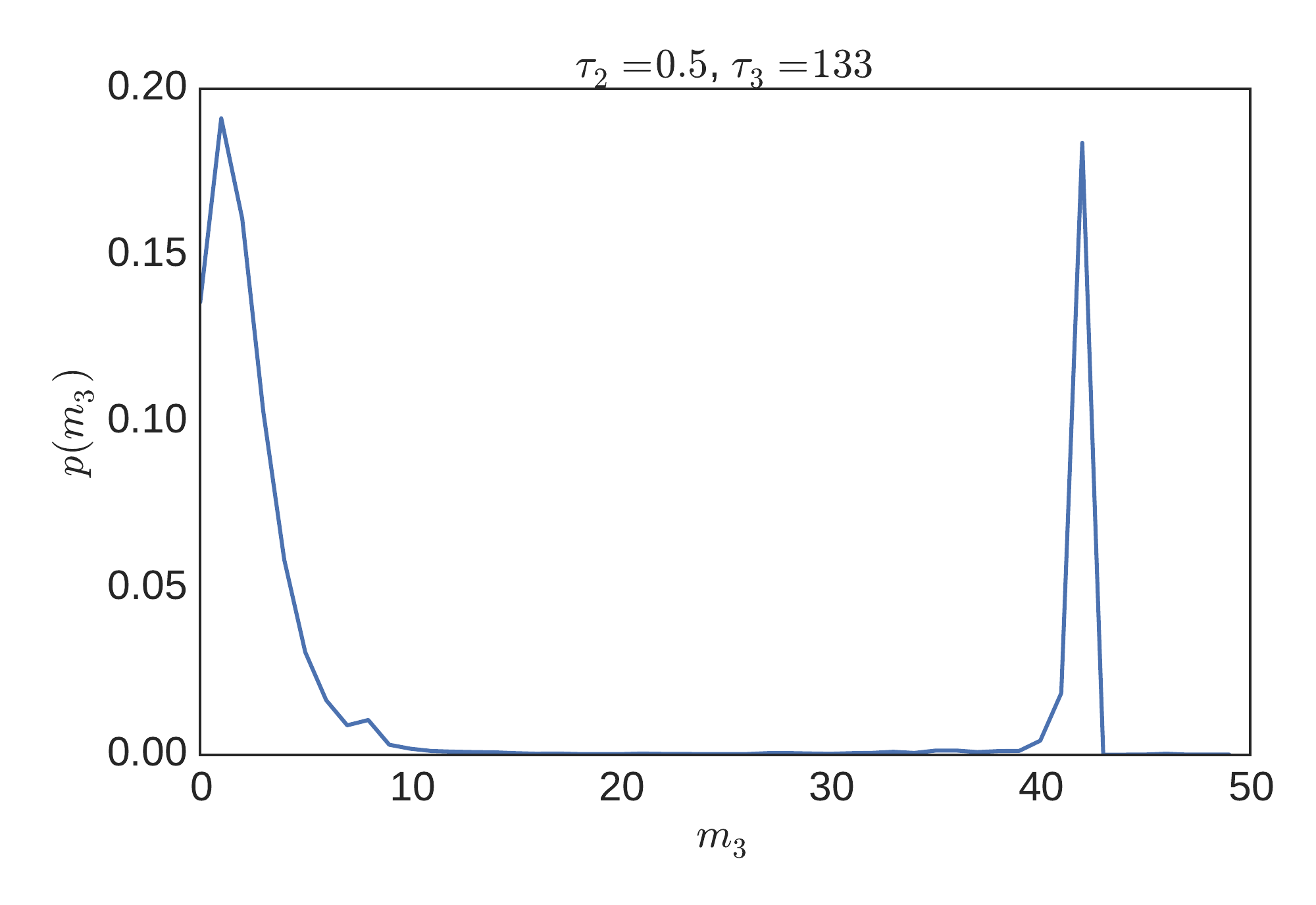}
\caption{The distribution of the number of triply visited sites $m_3$ is clearly bimodal at the point when $\tau_2$ and $\tau_3$ cross the line of suspected first-order transitions.%
}\label{IG-double-peak}
\end{center}
\end{figure}

Now we turn our attention to the generalisation of the VISAW model on the triangular lattice that explicitly includes three body interactions we call the IG model. We start in the same way by considering the density plot of the logarithm of the largest eigenvalue of the matrix of second derivatives of the free energy with respect to $\tau_2$ and $\tau_3$ in Figure~\ref{IG-hessian}.
The overall features of the diagram seem similar, barring a rescaling of the parameter values where the corresponding features appear. There are broadly three parts to the plot where the fluctuations are relatively small: one near the origin, one for large $\tau_3$ at small $\tau_2$, and finally one region for large $\tau_2$ at small $\tau_3$. This indicates three phases with phase boundaries between each meeting at a point close to $(\tau_2,\tau_3)=(4.5, 125)$. We looked for the location of the meeting point by considering the density plot of the ratio of the smallest to largest eigenvalue of the matrix of second derivatives of the free energy with respect to $\tau_2$ and $\tau_3$ in Figure~\ref{IG-hessian-ratio}.
If we set $\tau_3=1$ and plot the specific heat (Figure~\ref{IG-spec-heat-2}) we see little evidence of a transition although one can detect a weak transition on examining the third derivative (not shown here). This is exactly like the expected behaviour of the DS model.
On the other hand by fixing $\tau_2=0.5$ and varying $\tau_3$ a very strong transition is uncovered (see Figure~\ref{IG-spec-heat-3}).
If we consider the scaling of the maximum of the specific heat in a log-log plot (see  Figure~\ref{IG-spec-heat-diverge}), a super-linear divergence is found. This strong transition is already evident in the density plot of Figure~\ref{IG-hessian}.
To focus on the nature of this transition more closely we consider in Figure~\ref{IG-double-peak}  the distribution of triply-visited sites at the transition when $\tau_2=0.5$. A very clear bimodal distribution is seen once again allowing us to infer the existence of a first-order phase transition in the thermodynamic limit.

\begin{figure}[ht!]
\begin{center}
\includegraphics[width=\columnwidth]{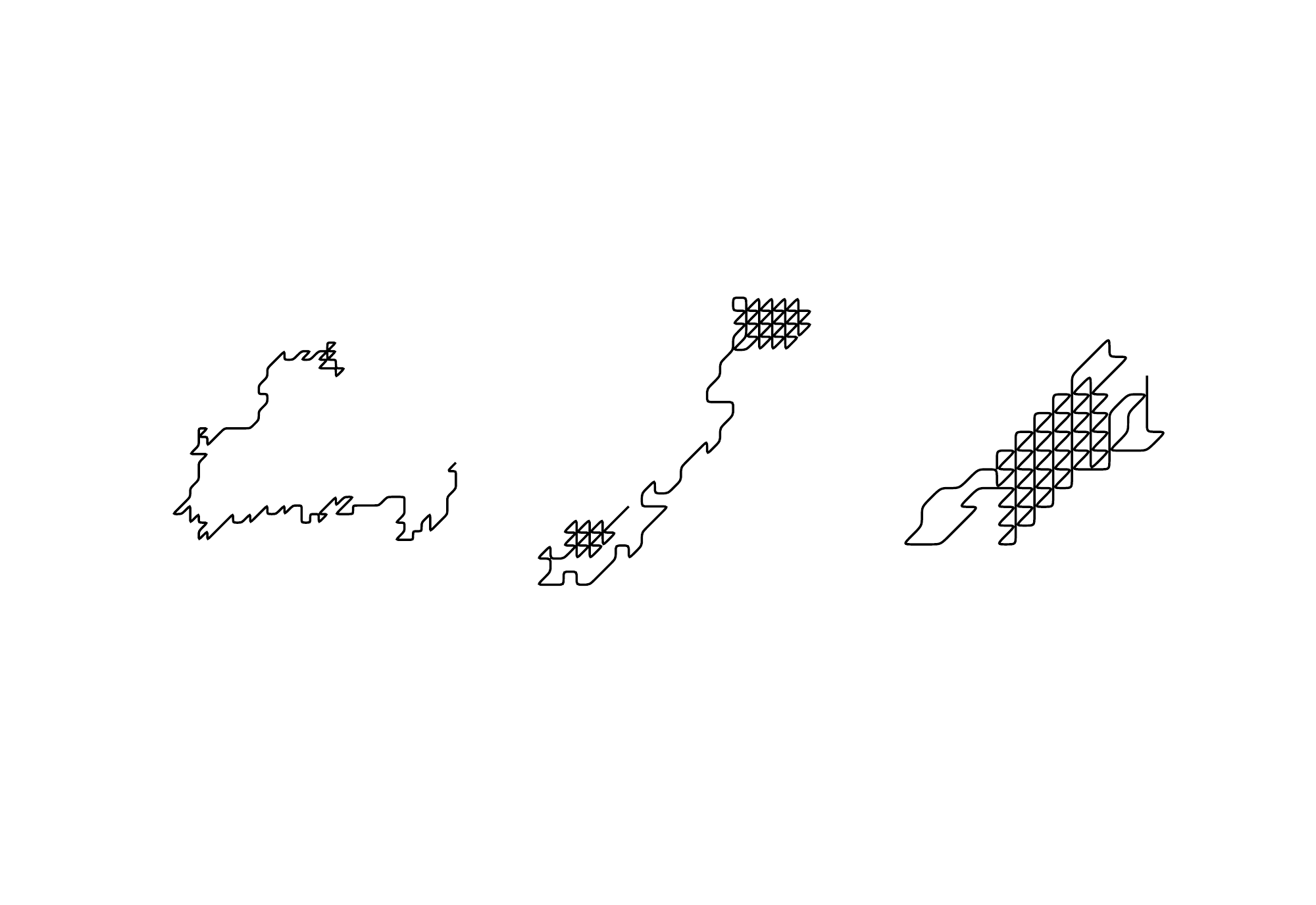}
\vspace{-18ex}
\caption{Typical VISAW configurations with $\tau_2 = 1$ with different values of $\tau_3$, from left to right: $\tau_3$ = 150, $\tau_3$ = 180, $\tau_3$ = 200. The configurations illustrate the co-existence of fully dense and swollen parts of the polymer, demonstrating the first-order nature of the transition.}\label{IG-typical-configs}
\end{center}
\end{figure}

In Figure~\ref{IG-typical-configs} we give typical configurations either side of, and at, the transition when $\tau=0.5$. 
The high temperature (low $\tau_2$) configuration is extended and would be indistinguishable from a self-avoiding walk at this scale, whereas for the low temperature point the configuration is made up of almost exclusive of triply visited sites. The configuration at the transition is made of extended and dense parts rather than being intermediate  in a non-trivial way. While a priori it is not expected that ordered will appear at low temperatures in this model, on closer inspection of the configurations they display clear  anisotropy indicating a crystal-like structure as we have seen in the honeycomb lattice model above.

\begin{figure}[ht!]
\begin{center}
\includegraphics[width=0.6\columnwidth]{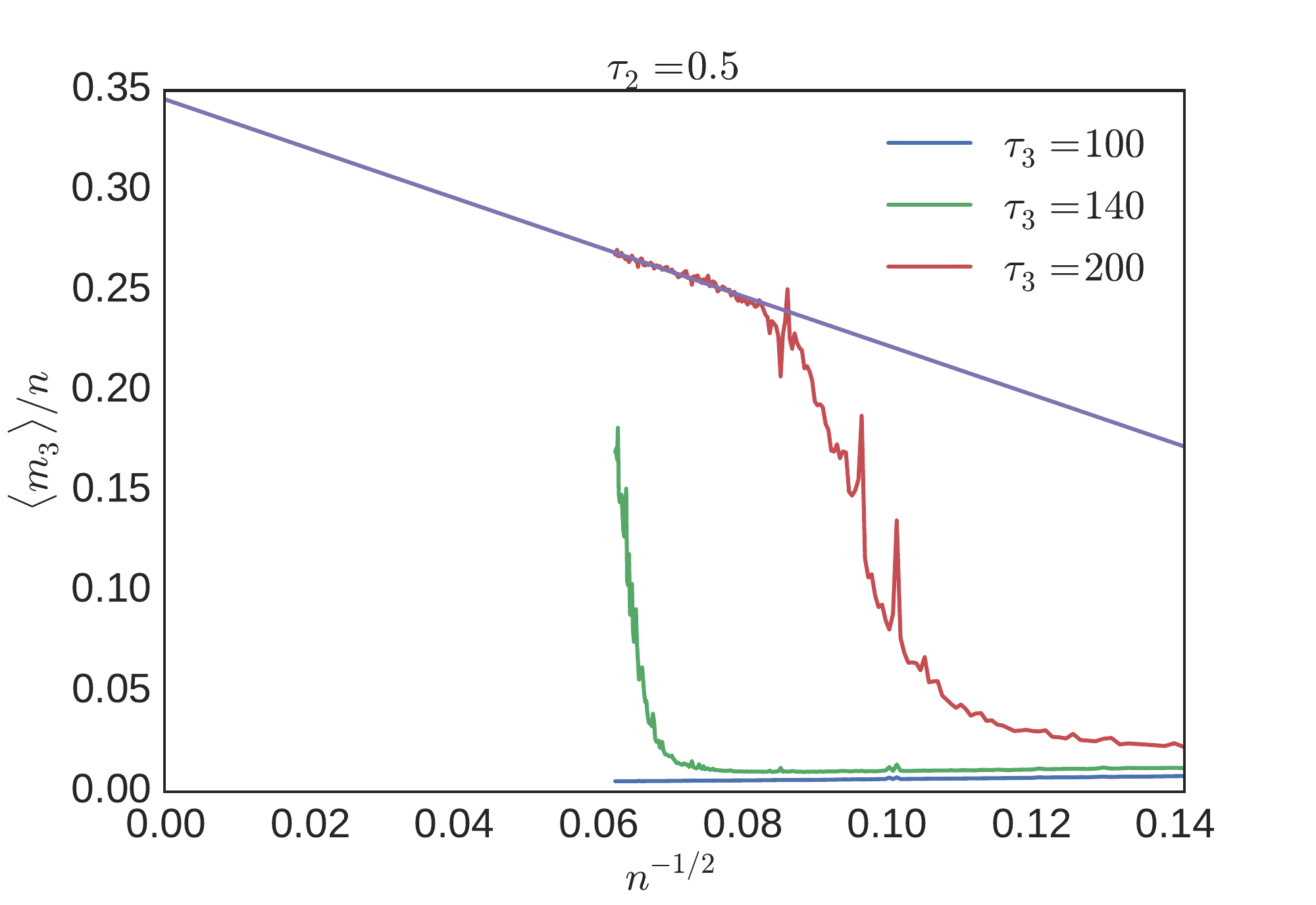}
\caption{Plot of  $\<m_3\>/n$ as a function of $n^{-1/2}$ for various values of $\tau_3$. In this plot $\tau_2$ is fixed to 0.5. The line is the graph of the linear extrapolation for $\tau_3 = 200$. For $\tau_3 = 200$ we estimate the $\lim_{n\rightarrow\infty} \<m_3\>/n$ to be close to $0.34$ --- a value of $1/3$, which is the maximum attainable asymptotic value, indicates a fully dense phase.%
}\label{IG-low-temp-order}
\end{center}
\end{figure}

If the high $\tau_3$ phase of the model when $\tau_2=0.5 $ indeed only contains triply-visited sites we would expect the density of triply visited sites to attain the value $1/3$ in the thermodynamic limit. In Figure~\ref{IG-low-temp-order} we plot the density of triply visited sites for various values of $\tau_3$ and for the largest value of $\tau_3=200$ the extrapolated value of the density is indeed close to $1/3$. While  the lengths we consider are relatively short for this analysis we are relatively confident in drawing this conclusion.

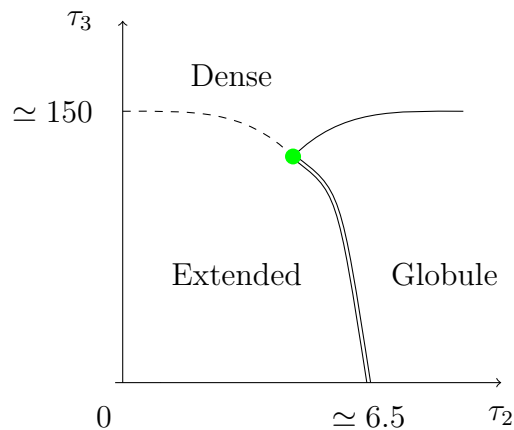
\begin{figure}[ht!]
\begin{center}
\begin{tikzpicture}[x=0.166cm,y=0.008cm,scale=3]
  \coordinate (A) at (6.5,0);
  \coordinate (C) at (4.5,125);
  \coordinate (B) at (0,150);

  \coordinate (C1) at ($(C)!0.25cm!10:(DS)$);

  \draw[double,double distance=1pt] (A) .. controls (C1) .. (C);

        \draw (1.0,60) node[align=left, right]{Extended};
   \draw (6.8,60) node[align=left, right]{Globule};
   \draw (1.5,170) node[align=left, right]{Dense}; 
   
         \draw[dashed] (0,150) to [out=0, in =135] (4.5,125);
    \draw[solid] (4.5,125) to [out=45, in =180] (9,150);
  \draw[->] (-0.2,0)
  	-- (0,0) node[left=7pt,below=5pt] {0}
    -- (A) node[below=5pt] {$\simeq 6.5$}
    -- (10,0) node[below=5pt] {$\tau_2$};
    
  \draw[->] (0,-0.2)
  	-- (B) node[left=7pt] {$\simeq 150$}
    -- (0,200) node[left=7pt] {$\tau_3$};

  \draw[black,solid] (1,-0.2) -- (1,0) ;

  \fill[green] (C) circle (1pt);
\end{tikzpicture}
\caption{Schematic phase diagram for generalised Interacting Grooves (IG) on the triangular lattice. There are three phases: Extended, Globule and Dense. The line of second-order $\theta$-like phase transitions between the Extended and Globule phases is marked with double lines while the first-order line between the Extended and Dense phases is marked with a dashed line. The boundary between the two low temperature phases of Globule and Dense, which we expect to be second-order, is marked with a single solid line. The meeting of these is marked by a bullet (green) at $(\omega_2,\omega_3)=(4.5,125)$.}\label{IG-phase-diagram}
\end{center}
\end{figure}

Piecing together the information we have at hand for our interacting groove model we provide a schematic phase diagram in Figure~\ref{IG-phase-diagram}.
Once again we have three phase boundaries meeting at a point marked with a bullet separating three phases marked as extended, globule and dense. There is no reason why the high $\tau_3$ phase will be ordered in any way, though it may be fully-dense, so we have been careful  not to label it as a crystal.

\section{Conclusion}
We have studied two models of two-dimensional polymer collapse based upon different lattice configurations that can touch or not, though not cross, in a larger parameter space where two and three body interactions are explicitly differentiated. In particular, we generalise the model studied by Duplantier and Saleur \cite{duplantier1987a-a} that describes the classical $\theta$-point universality class. We have shown using Monte Carlo simulation that this generalised model admits a first-order transition and a probable higher order multi-critical point in addition to the DS universality class. A similar scenario can be found in a generalisation of the VISAW model on the triangular lattice based upon configurations called grooves which are non-crossing trails. We conclude that the introduction of explicit three body interactions in either walk or groove based configurational polymer models lead to broadly similar phase diagrams that admit three phases and transitions as just described. The phase diagrams proposed here are broadly similar to that conjectured by Doukas \emph{et al.\ }\cite{doukas2010a-:a}  (see Figure~22) for interacting trails (ISAT) on the triangular lattice. So even though the three sets of underlying configurations of ISAW, ISAT and IG are all different in that walks, trails and grooves are either geometrically (walks and grooves) or topologically (grooves and trails)  different, they seem to give rise to broadly similar phase diagrams.

The remaining questions that arise are whether there are truly a low temperature transition from a globular phase at low three body interaction to a dense phase at larger values, and so the nature of that transition, whether the dense phase in the DS model which seems to be anisotropic is really similar to the triangular lattice groove model one, and finally the nature of the point separating the $\theta$-universality class from the first-order transition line in each phase diagram.

\ack

Financial support from the Australian Research Council via its support
for the Centre of Excellence for Mathematics and Statistics of Complex
Systems and through its Discovery program is gratefully acknowledged
by the authors. A L Owczarek thanks the School of Mathematical
Sciences, Queen Mary, University of London for hospitality.

\end{document}